\documentclass[preprint,pra,showpacs,nofootinbib]{revtex4}
\usepackage{mathrsfs}
\usepackage[mathscr]{euscript}
\usepackage{float,epsfig}
\usepackage{dcolumn}
\usepackage{bm}
\usepackage{graphicx}
\usepackage{dcolumn}
\usepackage{amsmath,amssymb,amsthm}
\usepackage[colorlinks=true,linkcolor=blue]{hyperref}
\usepackage{booktabs}

\newcommand{\bea}{\begin{eqnarray}}
\newcommand{\eea}{\end{eqnarray}}
\newcommand{\beq}{\begin{equation}}
\newcommand{\eeq}{\end{equation}}
\newcommand{\nn}{\nonumber}
\def\/{\over}
\newcommand{\bra}[1]{\langle#1|}
\newcommand{\ket}[1]{|#1\rangle}

\begin{document}
\title{\bf Quantum entanglement generation in de Sitter spacetime}
\author{ Jiawei Hu$^{1}$ and Hongwei Yu$^{1,2,}
$\footnote{Corresponding author} }
\affiliation{
$^1$ Institute of Physics and Key Laboratory of Low Dimensional Quantum Structures and Quantum Control of Ministry of Education,\\
Hunan Normal University, Changsha, Hunan 410081, China \\
$^2$ Center for Nonlinear Science and Department of Physics, Ningbo
University, Ningbo, Zhejiang 315211, China}

\begin{abstract}

We study, in the framework of open quantum systems, the entanglement generation between two mutually independent static two-level atoms in de Sitter spacetime. We treat the two-atom system as an open system in interaction with a bath of fluctuating conformally coupled massless scalar fields in the de Sitter invariant vacuum and derive the master equation which governs its evolution. With the help of the partial transposition criterion, we analyze entanglement generation between the two initially separable atoms and find that static atoms in de Sitter spacetime exhibit distinct characteristics from those in a thermal bath in flat spacetime in terms of entanglement generation. So, in, principle, one can tell whether he is in a thermal bath  or in de Sitter space by checking the entanglement creation between two initially separable static atoms in certain circumstances.

\end{abstract}

\maketitle

\section{Introduction}

Quantum entanglement describes a more intimate relationship between physical systems than one may encounter in the classical
world. A pure state of $n$ subsystems is entangled if it cannot be expressed as a direct product of the states of the subsystems, while for a mixed state, it is called entangled if it cannot be written as a convex combination of product states~\cite{Horodecki09}.  Quantum entanglement is at the heart of quantum information science~\cite{information}, since it  provides a key resource in many fascinating  applications, such as quantum communication~\cite{communication}, quantum teleportation~\cite{teleportation}, quantum cryptography~\cite{cryptography}, quantum metrology~\cite{metrology} and so on. One of the major barriers to the realization of the quantum information technologies is the environmental induced decoherence and dissipation to the quantum systems, which may cause entanglement degradation. However, an external environment can also provide indirect interactions between subsystems through correlations that exist. Therefore, there are certain situations where the environment may create entanglement rather than destroy it~\cite{ent-cre,Benatti-prl}. For two non-interacting atoms immersed in a common thermal bath, it has been found that, entanglement generation can be controlled by varying the bath temperature and the distance between the atoms, and the entanglement can persist even in the asymptotic equilibrium state for vanishing atom separation~\cite{Benatti-job}.

Recently, studies of entanglement generation caused by the correlations induced by the environment have been further carried out in many different situations including curved background~\cite{Benatti-pra,yu-prd-07,Martin-cqg-12,hu-jhep-11,steeg,Nambu}. 
 Benatti {\it et al.} have found that a uniformly accelerating two-atom system interacting with a bath of fluctuating scalar fields in the Minkowski vacuum will turn out to be entangled~\cite{Benatti-pra},  and this may be considered as a manifestation of the Unruh effect~\cite{Unruh}. Their work has been extended to the case with the presence of a reflecting boundary~\cite{yu-prd-07}, and it has been shown that, although the radiative properties for a single uniformly accelerated atom coupled to the scalar vacuum is the same as that immersed in a thermal bath, accelerated atoms may exhibit distinct features from static ones in terms of the entanglement creation~\cite{yu-pra-07}. Recently, a similar issue of quantum entanglement generation has been studied for inertial atoms in de Sitter spacetime~\cite{steeg},  where a pair of initially separable atoms is coupled to the scalar fields for a finite duration such that the atoms are kept causally independent, and then the negativity is examined for the final state. Although a single atom behaves as if there were a thermal bath at the Gibbons-Hawking temperature~\cite{GH-temp}, in certain circumstances, atoms in a thermal bath can get entangled while those in the corresponding de Sitter space can not~\cite{steeg}. Let us also note here that the classical and quantum correlations of scalar fields in de Sitter spacetime are investigated using a two-level detector model in interaction with both conformally coupled and minimally coupled massless scalar fields in Ref.~\cite{Nambu}.

In the present paper, we plan to study the entanglement generation for two mutually independent static two-level atoms in de Sitter spacetime in the framework of open quantum systems. Being the unique maximally symmetric curved spacetime, de Sitter spacetime is the simplest nontrivial curved background 
 and extensive studies have been done on the quantization of scalar fields in this spacetime
\cite{QFTCS,Tagirov,Bunch-Davies,Ford,Schomblond,Mottola,Allen-85,Allen-87,Polarski}. De Sitter spacetime is also of importance in cosmology since current cosmological observations in association with the inflation theory suggest that our universe may approach the exponentially expanding de Sitter phase both in the far past and the far future. Our interest in static atoms also lies in the fact that static atoms have an inherent acceleration in de Sitter spacetime. It is therefore worth comparing the entanglement generation for static atoms in de Sitter spacetime with that for the static ones in a thermal bath in flat spacetime.
Here we note that the approach applied in the current paper is different from that in Ref.~\cite{steeg,Nambu}, where the interactions between the atoms and fields are modulated by a window function such that the atom-field interactions can be considered as noncausal events, and the evolution of the atoms can be regarded as unitary except for a finite duration when the interaction is switched on. In the current paper, we treat the two-atom system as an open system which interacts with a bath of fluctuating vacuum scalar fields in de Sitter spacetime and which therefore evolves non-unitarily, and by tracing over the field degrees of freedom we derive the master equation which governs the atom's evolution. In so doing, we are able to examine the entanglement dynamics, in particular, the entanglement generation at the beginning of evolution and the entanglement remaining in the late asymptotic times.  At this point, let us note that the thermal nature in de Sitter spacetime has also been investigated through the radiative properties of atoms using the so-called DDC formalism~\cite{yu-jhep-08,yu-prd-10} and in the framework of open quantum systems~\cite{yu-prl}.


\section{Two Atom Master Equation}

The system we study is composed of two mutually independent static two-level atoms in interaction with a bath of conformally coupled massless scalar fields in vacuum in de Sitter spacetime. The total Hamiltonian takes the form
\begin{equation}
 H=H_A+H_F+H_I\;.
\end{equation}
Here $H_A$ is the Hamiltonian of the two-atom system,
\begin{equation}
H_A=H_A^{(1)}+H_A^{(2)},\quad
H_A^{(\alpha)}={\omega\over 2}\,\sigma_3^{(\alpha)},~~(\alpha=1,2),
\end{equation}
where $\sigma_i^{(1)}=\sigma_i\otimes{\sigma_0}$, $\sigma_i^{(2)}={\sigma_0}\otimes\sigma_i$, with $\sigma_i~(i=1,2,3)$ being the Pauli matrices, $\sigma_0$ the $2\times2$ unit matrix, and $\omega$ the energy level spacing of the atoms. $H_F$ is the Hamiltonian of the massless scalar fields in de Sitter spacetime, details of which do not need to be specified here. $H_I$ is the Hamiltonian describing the interaction between the atoms and the scalar fields, and we suppose it to be in the following form in analogy to the electric dipole interaction~\cite{Benatti-pra}
\begin{equation}
H_I=\mu\,[(\sigma_{2} \otimes \sigma_0)\Phi(t,{\bf x}_1)
   +(\sigma_{0} \otimes \sigma_{2})\Phi(t, {\bf x}_2)\,]\;,
\end{equation}
where $\mu$ is the coupling constant which is assumed to be small.

The time evolution of the total system in the proper time $\tau$ is governed by the von Neumann equation $\partial_\tau\rho_{\text{tot}}(\tau)
=-i[H,\rho_{\text{tot}}(\tau)]$,
with the initial state taking the form $\rho_{\text{tot}}(0)=\rho(0)\otimes\ket{0}\bra{0}$,
where $\ket{0}$ is the vacuum state of the scalar fields and $\rho(0)$ describes the initial state of the two-atom system. The reduced dynamics of the two-atom system is derived by tracing over the field degrees of freedom $\rho(\tau)=\text{Tr}_F[\rho_{\text{tot}}(\tau)]$, and it takes the Kossakowski-Lindblad form~\cite{Lindblad,open} in the limit of weak-coupling
\begin{equation}\label{master}
{\partial\rho(\tau)\/\partial\tau}
=-i\big[H_{\rm eff},\,\rho(\tau)\big]+{\cal L}[\rho(\tau)]\ ,
\end{equation}
with
\begin{equation}
H_{\rm eff}=H_A-\frac{i}{2}\sum_{\alpha,\beta=1}^2\sum_{i,j=1}^3
H_{ij}^{(\alpha\beta)}\,\sigma_i^{(\alpha)}\,\sigma_j^{(\beta)}\ ,
\end{equation}
and
\begin{equation}
{\cal L}[\rho]
={1\over2} \sum_{\alpha,\beta=1}^2\sum_{i,j=1}^3
 C_{ij}^{(\alpha\beta)}
 \big[2\,\sigma_j^{(\beta)}\rho\,\sigma_i^{(\alpha)}
 -\sigma_i^{(\alpha)}\sigma_j^{(\beta)}\, \rho
 -\rho\,\sigma_i^{(\alpha)}\sigma_j^{(\beta)}\big]\ .
\end{equation}
The elements of the matrices $C_{ij}^{(\alpha\beta)}$ and
$H_{ij}^{(\alpha\beta)}$ are determined by the Fourier and Hilbert transforms of the following field correlation functions
\begin{equation}\label{green}
\mathrm{}G^{(\alpha\beta)}(\tau-\tau')
=\langle\Phi(\tau,\mathbf{x}_{\alpha})\Phi(\tau',\mathbf{x}_\beta)
 \rangle\;,
\end{equation}
${\cal G}^{(\alpha\beta)}(\lambda)$ and  ${\cal K}^{(\alpha\beta)}(\lambda)$, which are defined  respectively as
\begin{equation}\label{fourierG}
{\cal G}^{(\alpha\beta)}(\lambda)
=\int_{-\infty}^{\infty} d\Delta\tau\,
 e^{i{\lambda}\Delta\tau}\, G^{(\alpha\beta)}(\Delta\tau)\; ,
\end{equation}
\begin{equation}
{\cal K}^{(\alpha\beta)}(\lambda)
=\frac{P}{\pi i}\int_{-\infty}^{\infty} d\omega\
 \frac{{\cal G}^{(\alpha\beta)}(\omega)}{\omega-\lambda} \;,
\end{equation}
where $P$ stands for principal value. Then the Kossakowski matrix $C_{ij}^{(\alpha\beta)}$ can be written explicitly as
\beq
C_{ij}^{(\alpha\beta)}
= A^{(\alpha\beta)}\delta_{ij}
 -iB^{(\alpha\beta)}\epsilon_{ijk}\,\delta_{3k}
 -A^{(\alpha\beta)}\delta_{3i}\,\delta_{3j}\;,
\eeq
where
\begin{equation}\label{abc1}
\begin{aligned}
A^{(\alpha\beta)}
={\mu^2\/4}\,[\,{\cal G}^{(\alpha\beta)}(\omega)
 +{\cal G}^{(\alpha\beta)}(-\omega)]\;,\\
B^{(\alpha\beta)}
={\mu^2\/4}\,[\,{\cal G}^{(\alpha\beta)}(\omega)
 -{\cal G}^{(\alpha\beta)}(-\omega)]\;.
\end{aligned}
\end{equation}
 $H^{(\alpha\beta)}_{ij}$
can be similarly obtained by replacing ${\cal
G}^{(\alpha\beta)}$ with ${\cal
K}^{(\alpha\beta)}$ in the above equations.

\section{The condition for entanglement creation}

Four dimensional de Sitter spacetime can be viewed as the hyperboloid
\beq
z_0^2-z_1^2-z_2^2-z_3^2-z_4^2=-\alpha^2\;,
\eeq
embedded in five dimensional Minkowski spacetime
\beq
ds^2=dz_0^2-dz_1^2-dz_2^2-dz_3^2-dz_4^2\;.
\eeq
By applying the following parametrization
\beq
\begin{aligned}
&z_0=\sqrt{\alpha^2-r^2}\sinh{t/\alpha}\;,\\
&z_1=\sqrt{\alpha^2-r^2}\cosh{t/\alpha}\;,\\
&z_2=r\cos\theta\;,\\
&z_3=r\sin\theta\cos\phi\;,\\
&z_4=r\sin\theta\sin\phi\;,
\end{aligned}
\eeq
the static de Sitter metric is obtained
\beq
ds^2=\biggl(1-\frac{r^2}{\alpha^2}\biggr)d{t}^2
    -\biggl(1-\frac{r^2}{\alpha^2}\biggr)^{-1}dr^2
    -r^2(d\theta^2+\sin^2\theta d\phi^2)\;,
\eeq
where $\alpha=\sqrt{3/\Lambda}$ with $\Lambda$ being the cosmological constant. The origin of the coordinates $r=0$ corresponds to the position of the observer, so the coordinate singularity at $r=\alpha$ is the cosmological horizon for the observer, and the coordinates $(t,~r,~\theta,~\phi)$  cover only part of the de Sitter spacetime. Here, for simplicity, we assume that two static atoms are held at the same distance to the observer at $r=0$ but at  different azimuthal angles, i.e., they are assume to be located respectively at $(t,~r,~\theta,~\phi)$ and $(t,~r,~\theta',~\phi)$.

In dealing with curved spacetime, a delicate issue is how to determine the vacuum state of the quantum fields. Here we choose the de Sitter-invariant vacuum state, since it preserves the de Sitter invariance and is considered to be a natural choice of vacuum state in this spacetime~\cite{Allen-85}. The Wightman function of a conformally coupled massless scalar field in the de Sitter invariant vacuum takes the form~\cite{QFTCS,Tagirov}
\beq
G^+(x,x')=-\frac{1}{4\pi^2}
           \frac{1}{(z_0-z_0')^2-(z_1-z_1')^2-(z_2-z_2')^2
           -(z_3-z_3')^2-(z_4-z_4')^2}\;.
\eeq
Then the correlation functions for two spacetime points where the atoms are located can be easily derived as
\beq
G^{(11)}(x,x')=G^{(22)}(x,x')
=-\frac{1}{16\pi^2\kappa^2\sinh^2(\frac{\tau-\tau'}{2\kappa}-i\epsilon)}\;,
\eeq
\beq
G^{(12)}(x,x')=G^{(21)}(x,x')
=-\frac{1}{16\pi^2\kappa^2}
  \frac{1}{\sinh^2(\frac{\tau-\tau'}{2\kappa}-i\epsilon)
  -{r^2\/\kappa^2}\sin^2{\Delta\theta\/2}}\;,
\eeq
with $\kappa=\sqrt{g_{00}}\;\alpha$\;, where we have used the relation $\Delta\tau=\sqrt{g_{00}}\;\Delta{t}$.
The Fourier transforms with respect to the proper time are
\begin{eqnarray}
&&{\cal G}^{(11)}(\lambda)={\cal G}^{(22)}(\lambda)
=\frac{1}{2\pi}
 \frac{\lambda}{1-e^{-2\pi\kappa\lambda}},\\
&&{\cal G}^{(12)}(\lambda)={\cal G}^{(21)}(\lambda)
=\frac{1}{2\pi}
 \frac{\lambda}{1-e^{-2\pi\kappa\lambda}}f(\lambda,L/2)\;,
\end{eqnarray}
where $f(\lambda,z)$ is defined as
\begin{equation}
f(\lambda,z)
=\frac{\sin[{2\kappa\lambda}\sinh^{-1}(z/\kappa)]}
 {2z\lambda\sqrt{1+z^2/\kappa^2}}\;,
\end{equation}
and $L=2r\sin({\Delta\theta/2})$ is the usual Euclidean distance between the two points $(r,~\theta,~\phi)$ and $(r,~\theta',~\phi)$. Plugging the Fourier transforms into Eq.~(\ref{abc1}), one obtains
\begin{eqnarray}
&&C_{ij}^{(11)}=C_{ij}^{(22)}
=A_1\,\delta_{ij}-iB_1\epsilon_{ijk}\,\delta_{3k}
 -A_1\delta_{3i}\,\delta_{3j}\;,\\
&&C_{ij}^{(12)}=C_{ij}^{(21)}
=A_2\,\delta_{ij}-iB_2\epsilon_{ijk}\,\delta_{3k}
 -A_2\delta_{3i}\,\delta_{3j}\;,
\end{eqnarray}
where
\begin{eqnarray}
\begin{aligned}
&A_1={\mu^2\omega\over8\pi}\,
     \frac{e^{2\pi\kappa{\omega}}+1}{e^{2\pi\kappa{\omega}}-1}\;,\\
&A_2={\mu^2\omega\over8\pi}\,
     \frac{e^{2\pi\kappa{\omega}}+1}{e^{2\pi\kappa{\omega}}-1}\,f(\omega,L/2)\;,\\
&B_1={\mu^2\omega\over8\pi}\;,\\
&B_2={\mu^2\omega\over8\pi}\,f(\omega,L/2)\;.
\end{aligned}
\label{abc2}
\end{eqnarray}

Now we examine whether two initially separable atoms can get entangled at the beginning of evolution. For this purpose, the partial transposition criterion~\cite{ppt} can be applied, i.e., a two-atom state $\rho$ is entangled if and only if the operation of partial transposition of $\rho$ does not preserve its positivity. Here we focus our study on pure initial states since that if the bath can not entangle these states, it can  neither entangle their mixtures. Further, for simplicity, the initial state of the two atoms is supposed to be $\rho(0)=|+\rangle\langle+|\otimes|-\rangle\langle-|$. In this case, the partial transposition criterion leads to a more manageable condition, i.e., entanglement can be generated at the neighborhood of time $t=0$ if and only if~{\cite{Benatti-prl,Benatti-job}
\begin{equation}\label{condition}
\bra{u}C^{(11)}\ket{u} \bra{v}(C^{(22)})^T)\ket{v}
< |\bra{u}{\rm Re}(C^{(12)})\ket{v}|^2\;,
\end{equation}
in which the superscript $T$ denotes matrix transposition, and the three-dimensional vectors $\ket{u}$ and $\ket{v}$ take the form as $u_i=v_i=\{1,-i,0\}$.  The condition (\ref{condition}) becomes after some straight calculations using Eq.~(\ref{abc1}),
\begin{equation}\label{condition1}
\bigg({A_2\/A_1}\bigg)^2+\bigg({B_1\/A_1}\bigg)^2>1\;,
\end{equation}
where
\beq\label{a2a1}
\bigg({A_2\/A_1}\bigg)^2
=\frac{\sin^2[{2\kappa\,\omega}\sinh^{-1}(L/2\kappa)]}
 {\omega^2L^2(1+L^2/4\kappa^2)}\;,
\eeq
\beq\label{b1a1}
\bigg({B_1\/A_1}\bigg)^2
=\bigg(\frac{e^{2\pi\kappa{\omega}}-1}
       {e^{2\pi\kappa{\omega}}+1}\bigg)^2\;.
\eeq

Here both $A_2^2/A_1^2$ and $B_1^2/A_1^2$ range from 0 to 1. For a given atom, $B_1^2/A_1^2$ depends only on $\kappa$ or the effective temperature $T_{\rm eff}={1\over2\pi\kappa}$ which can be rewritten in the form
\beq
T_{\rm eff}^2=\bigg({1\/2\pi\alpha}\bigg)^2+\bigg({a\/2\pi}\bigg)^2
   =T_{\rm ds}^2+T_U^2\;,
\eeq
where $T_{\rm ds}={1\over2\pi\alpha}$ is the temperature of thermal radiation in de Sitter spacetime as perceived by a freely falling observer, and $T_{U}={a\over2\pi}$ is the Unruh temperature with
\beq
a={r\/\alpha^2}\bigg(1-{r\/\alpha^2}\bigg)^{-1/2}
 ={r\/\alpha}\,\kappa^{-1}\;
\eeq
being the inherent acceleration of a static atom in de Sitter spacetime. Therefore, $T_{\rm eff}$ actually arises as a combined effect of both the thermal nature of de Sitter spacetime characterized by the Gibbons-Hawking temperature $T_{\rm ds}$ and the Unruh effect related to the inherent proper acceleration $a$ of the static atoms. On the other hand, $A_2^2/A_1^2$ depends both on $\kappa$ and the separation $L$. For a given $\kappa$, $A_2^2/A_1^2$ oscillates as $L$ increases with a decreasing amplitude, from one when $L=0$ to z ero when $L$ goes to infinity. While for a given $L$, the relation between $A_2^2/A_1^2$ and $\kappa$ is somewhat complicated and depends on the specific value of $L$.  When the atoms are located near $r=0$, $\kappa$ will be very large since the cosmological constant is very small, so the effective temperature will be vanishingly small ($\sim10^{-30}$ K), and $B_1^2/A_1^2$ will approach to 1. At the same time, $A_2^2/A_1^2$ also approaches to 1 since $L\rightarrow 0$, so the inequality (\ref{condition1}) is satisfied. When the differences between the polar angles $\Delta\theta=0$, the separation of the two atoms $L$ is vanishing, then $A_2^2/A_1^2=1$, and the condition is also fulfilled unless the atoms are placed at the horizon where the effective temperature is infinitely high so that $B_1^2/A_1^2=0$. Thus for a given position $r$, there always exists an angle $\Delta\theta$ less than which entanglement between the two atoms can be generated, except for $r=\alpha$.

A comparison between the equations above and those for entanglement creation for two static atoms in a thermal bath in flat spacetime~\cite{Benatti-job} shows that static atoms in de Sitter spacetime behave differently from the ones that immersed in a thermal bath in flat spacetime in terms of entanglement generation at the beginning of evolution. To be specific, for a given atom, $A_2^2/A_1^2$ here depends both on the separation $L$ and the effective temperature, while the corresponding term for a thermal bath, i.e.,  Eq.~(37) in Ref.~\cite{Benatti-job}, depends on the separation $L$ only. This suggests that although a single static atom coupled to the fluctuating scalar fields in de Sitter invariant vacuum behaves the same as if it were in a thermal bath in a flat spacetime at a temperature $T_{\rm eff}$, the two situations can be distinguished by examining whether  entanglement is generated between  two initially separable static atoms. So, in principle, quantum entanglement creation can tell whether you are in a thermal bath in a flat spacetime  or in a de Sitter spacetime.

\section{Discussion and Conclusion}

In summary, we have studied the entanglement generation between two mutually independent static two-level atoms in de Sitter spacetime in the framework of open quantum systems. The two-atom system is treated as an open quantum system in interaction with a bath of fluctuating conformally coupled massless scalar fields in the de Sitter invariant vacuum, and the master equation describing its evolution is derived. We then analyze entanglement generation between the two initially separable atoms with the help of the partial transposition criterion, and find that the conditions for entanglement generation for static atoms in de Sitter spacetime are different from those in a thermal bath in the flat spacetime. So, in principle, one can tell whether he is in a thermal bath  or in de Sitter space by checking the entanglement creation between two initially separable static atoms in certain circumstances.

Now a comment is in order. We focus, in this paper, on the condition of entanglement generation for initially separable static atoms at the beginning of evolution with the help of the partial transposition criterion. However, this criterion can not tell us how the entanglement evolves and whether it can persist over time. Although the effects of decoherence and dissipation due to the interaction with environment will generally cause the initially entangled pairs to be separable, there are also cases in which the entanglement are preserved at late times. In order to quantify the late time entanglement of the two-atom system, we may take concurrence~\cite{concurrence} as a measurement. Following the same procedures as in Ref.~\cite{Benatti-job,Benatti-pra}, one finds that the asymptotic state concurrence is nonzero only if the distance between the atoms is vanishing, and it takes the same form as that of the static two-atom system in a thermal bath with a temperature $T_{\rm eff}$. This is not surprising, since the two atoms are at the same space point and the unique field correlations due to the non-trivial de Sitter geometry is absent.

\begin{acknowledgments}
This work was supported in part by the National Natural Science Foundation of China under Grants No. 11075083, No. 10935013 and No. 11375092; the Zhejiang Provincial Natural Science Foundation of China under Grant No. Z6100077;  the National Basic Research Program of China under Grant No. 2010CB832803; the Program for Changjiang Scholars and Innovative Research Team in University (PCSIRT,  No. IRT0964); the Hunan Provincial Natural Science Foundation of China under Grant No. 11JJ7001; the SRFDP under Grant No. 20124306110001 and Hunan Provincial Innovation Foundation For Postgraduate under Grant No. CX2012A009.
\end{acknowledgments}

\end{document}